# Late Breaking Results: Breaking Symmetry— Unconventional Placement of Analog Circuits using Multi-Level Multi-Agent Reinforcement Learning

Supriyo Maji, Linran Zhao, Souradip Poddar, David Z. Pan
*ECE Department, The University of Texas at Austin, Austin, TX, USA*
{smaji@alumni.purdue, {lrzhao@, souradippddr1@, dpan@ece.}utexas}.edu

*Abstract*— **Layout-dependent effects (LDEs) significantly impact analog circuit performance. Traditionally, designers have relied on symmetric placement of circuit components to mitigate variations caused by LDEs. However, due to non-linear nature of these effects, conventional methods often fall short. We propose an objective-driven, multi-level, multi-agent Q-learning framework to explore unconventional design space of analog layout, opening new avenues for optimizing analog circuit performance. Our approach achieves better variation performance than the state-of-the-art layout techniques. Notably, this is the first application of multi-agent RL in analog layout automation. The proposed approach is compared with non-ML approach based on simulated annealing.**

## I. INTRODUCTION

Analog circuit performance is affected by two types of variations: random and systematic. Random variations can be reduced by increasing device size, while systematic variations arising from layout-dependent effects (LDEs), require careful manual layout, which is time-consuming. Traditionally, designers use symmetric layout to improve variation tolerance, but this approach works only if circuit performance depends linearly on device parameter that exhibits linear variation [1]. When circuit performance depends non-linearly on device parameter or the device parameter has non-linear variation, a symmetric layout may not be effective [1][2]. Along with symmetry, another commonly used method for minimizing LDEs is putting dummies around. This, however, can double circuit area and introduce additional parasitics. Moreover, even with dummies included in a perfectly symmetric layout, non-linear variations may not cancel. This underscores the need for an objective-driven placement to address LDEs.

Recently, reinforcement learning (RL), a branch of machine learning, has emerged as a promising approach for addressing analog circuit optimization problems [3][4]. Among various RL algorithms, we adapt Q-learning for its suitability in addressing our problem. We leverage analog circuits' modularity to break down the placement optimization space into smaller subspaces and apply multi-agent RL, an advancing field within RL [5]. To the best of our knowledge, this is the first use of multi-agent RL in analog layout automation. We further enhance the technique by introducing two levels in Q-learning. Our multi-level, multi-agent RL approach is scalable. We compare with non-ML approach based on simulated annealing which has been extensively used in physical design [4]. The proposed framework generates unconventional, non-intuitive layouts that are better than the traditional symmetric style of layout [6][7][8].

## II. Analog Placement Optimization

Layout-dependent effects (LDEs) are critical considerations in analog placement, as they can alter device characteristics such as mobility, threshold voltage, and parasitics significantly [10]. This can introduce mismatch between circuit elements, significantly impacting overall circuit performance. For placing analog components, designers typically follow a signal flow graph representation [7]. To enhance matching performance, analog components are placed symmetrically, with sensitive transistors grouped according to primitives (e.g., input pair, load pair, current mirror etc) [6][9]. Fig. 1 illustrates two common layout strategies for analog circuits. Fig. 1(b) considers symmetry along Y-axis, while Fig. 1(c) considers symmetry along X- and Y-axes while following the groups shown in Fig. 1(a). Each method has its strengths and limitations. The first placement is easy to route but only compensates for variations along the Y-axis. The second method mitigates variations along X- and Y- axis but is difficult to route and may increase capacitance, which is undesirable in high-frequency applications. Expert designers use combination of these strategies during circuit placement. However, these layout styles are effective only when performance is linearly dependent on linearly varying parameters—a condition that does not generally hold true.

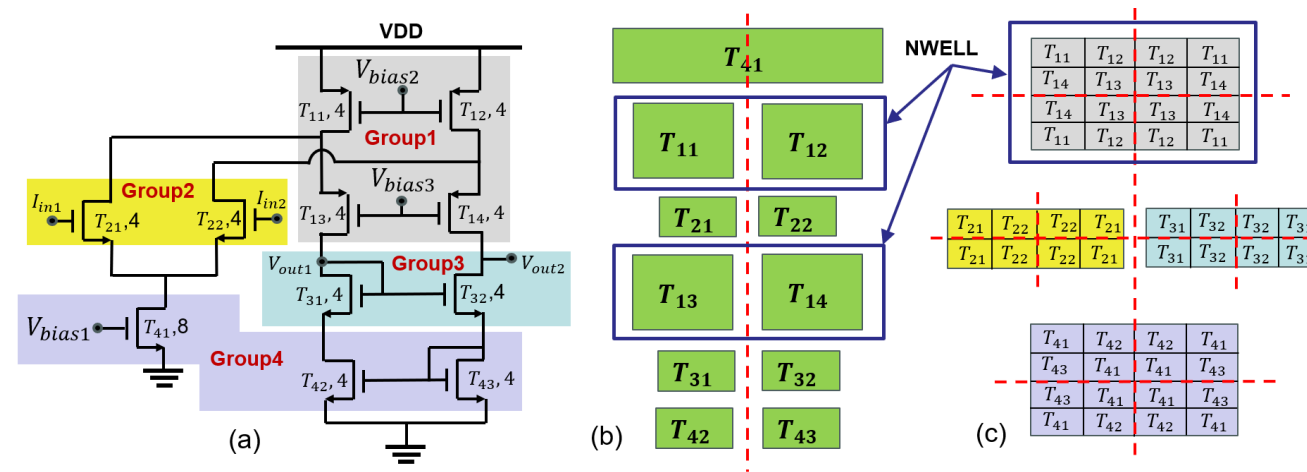

Fig. 1: (a) A folded-cascode OTA. Layout following (b) Y-axis symmetry [7][8], and (c) X- and Y- axis symmetry and grouping [6][9], as shown in (a).

### A. Multi-Level, Multi-Agent Q-learning for Circuit Placement

We follow a typical RL framework for placement optimization, where an agent interacts with a layout environment at discrete time steps. At each time step $t$, the agent observes the placement state $S_t$ and receives a reward $R_t$ reflecting the placement quality. It then chooses an action $A_t$ from the set of available actions to reposition a device resulting in a new state $S_{t+1}$, associated with a transition ($S_t$, $A_t$, $S_{t+1}$). The agent's objective is to explore various trajectories from the initial to a final placement, maximizing cumulative rewards over time until it converges on a placement solution that meets the target. In the example in Fig. 2(a), we show layout environment of a circuit. Fig. 2(b) shows five legal moves out of eight possible moves available for a unit device. During optimization, all units within a group remain connected. In the objective-driven placement, the quality of a move is checked with the simulator, guiding the algorithm toward better placement. In Q-learning, a Q-table, updated following the Bellman equation below, captures the state-action trajectories.

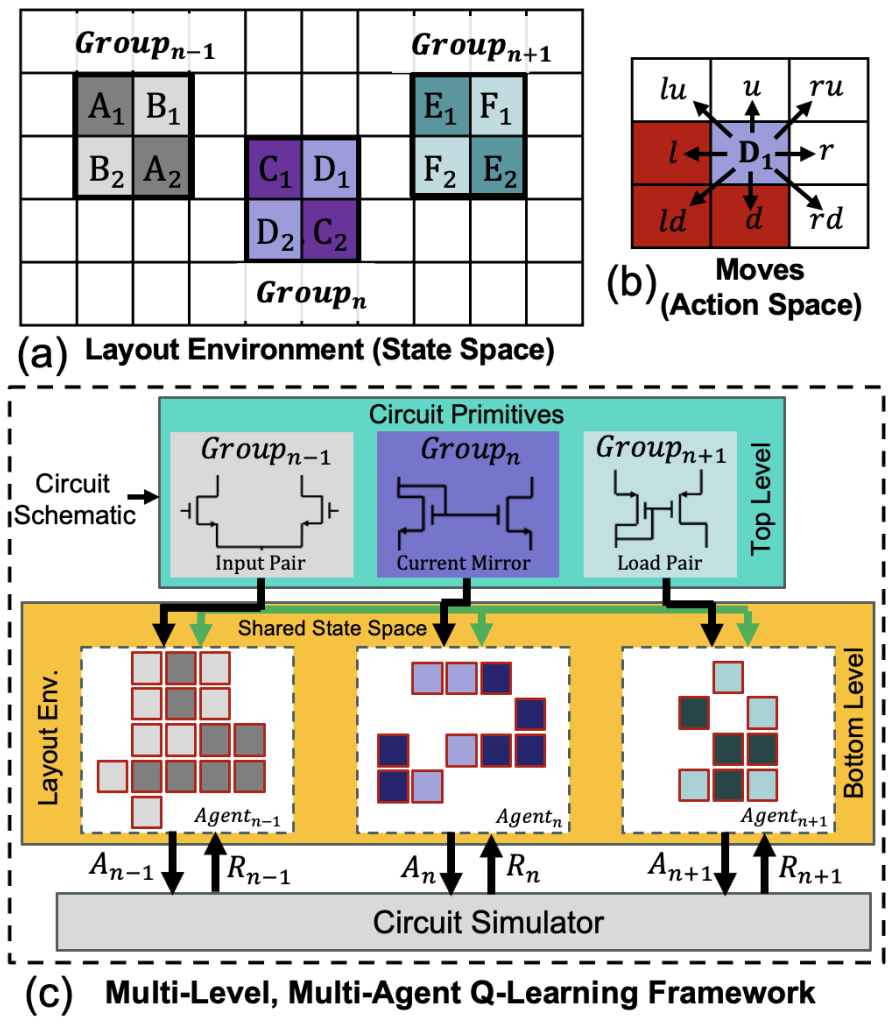

Fig. 2: (a) Layout environment showing three groups with two devices in each. Each device has two units. (b) Out of eight possible moves (action space), five moves are legal. (c) The proposed framework.

$$Q(S_t, A_t) \leftarrow (1-\alpha)Q(S_t, A_t) + \alpha[R_{t+1} + \gamma V(S_{t+1})] \quad (1)$$

$$V(s) \leftarrow \max_{a \in A} Q(s, a) \quad (2)$$

Here $\alpha$ is the learning rate and $\gamma$ is the discount factor. Q-table allows the algorithm to improve its decision-making over time. However, a key challenge lies in handling the growth of the Q-table as the number of devices increases. To address the scalability issue, we introduce a multi-level, multi-agent approach. At the top level, a Q-table is used to learn the movement of the groups, while at the bottom level, each agent maintains its own Q-table to learn the movement of units within a group. This hierarchical structure aligns with the grouping strategy in analog placement, where each agent represents a group in a multi-agent settings. Q-table updates are performed in an interleaved manner, ensuring conflict-free movement between agents. Fig. 2(c) shows the proposed framework.

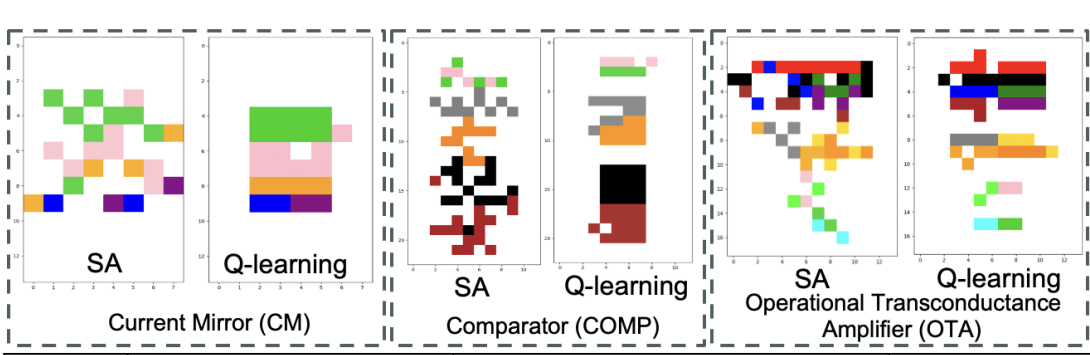

| Circuit | Mismatch (µA) | | | $FOM_{CM}$ (1/µA.µm$^2$) * | | | # Simulations | |
|---|---|---|---|---|---|---|---|---|
| | SOTA | SA | Q-Learning | SOTA | SA | Q-Learning | SA | Q-Learning |
| CM | 100 | **0.11** | 0.21 | 0.0003 | 0.09 | **0.1** | 345 | **50** |
| | Offset (mV) | | | $FOM_{COMP}$ (1/mV. µs. mW.µm$^2$) ** | | | | |
| COMP | 1.8 | 0.4 | **0.2** | 0.002 | 0.007 | **0.02** | 273 | **189** |
| | Offset (mV) | | | $FOM_{OTA}$ (KHz.deg/mV.µW.µm$^2$) *** | | | | |
| OTA | 23.6 | 14.6 | **13.6** | 1.05 | 1.13 | **1.75** | 216 | **88** |

* $FOM_{CM} = \frac{1}{\text{Mismatch.Area}}$ ** $FOM_{COMP} = \frac{1}{\text{Offset.Delay.Power.Area}}$ *** $FOM_{OTA} = \frac{\text{Gain.BW.PM}}{\text{Offset.Power.Area}}$

Fig. 3: Placement results. Color indicates different devices. For comparison, we show static mismatch/offset, FOM and # simulations required by the algorithms (best results in bold).

## III. Experimental Results

We implemented Q-learning and Simulated Annealing (SA) algorithms for circuit placement optimization using Python and SKILL. Automatic routing and post-layout extraction were performed using Virtuoso and Calibre. While routing was not optimized, its effects were included in the simulation. For the initial placement, we used signal flow graph to find relative placement location of the groups. Units within a group were placed sequentially. We evaluated the algorithms on a medium-sized Current Mirror (CM) and two large circuits: a Comparator (COMP) and an Operational Transconductance Amplifier (OTA) where mismatch/offset is critical. We set target mismatch/offset based on the best layout generated by state-of-the-art (SOTA) commercial and academic tools [10][6][7] using TSMC 40 nm technology. As shown in Fig. 3, unconventional layout had significantly better mismatch/offset performance than symmetric layout across all examples. We also reported a Figure of Merit (FOM) covering key metrics: CM (Mismatch, Area), COMP (Offset, Delay, Power, Area), and OTA (Gain, BW, PM, Offset, Power, Area). There was significant improvement in FOM across all examples. Note that while metrics such as Gain, BW, Delay, PM, Power, and Area can be optimized by transistor sizing, mismatch/offset due to LDEs cannot be predicted or mitigated at the schematic stage. We conclude that a symmetric layout is not optimal due to non-linearity in variations.

In the objective-driven placement, Q-learning was faster and produced solutions with better mismatch/offset, and FOM, compared to SA. What sets Q-learning apart from SA is its ability to learn and improve over time by gradually refining its policy. In contrast, SA does not involve learning; instead, it focuses on exploring solutions near the current best. This can make SA struggle to explore different trajectories. Q-learning can also be inefficient when optimal solution is far from the initial state or when the state-action space is large. In such cases, the dominance of exploration in Q-learning may prevent efficient exploitation of past learning. However, Q-learning proves to be effective here, since it is implemented in a multi-level, multi-agent settings.

## IV. Conclusions

We have proposed an objective-driven, simulation-based framework for analog circuit placement. The framework is technology and design agnostic and can be extended to handle analog/mixed-signal system layout. Our work will encourage the community to reassess the relevance of symmetric layout and explore the potential of multi-agent RL in addressing other problems in circuit design.